\documentclass[usenatbib]{mn2e}
\usepackage{natbib,epsfig,subfigure,lscape,longtable}

\def\mb{\hbox{{\rm M}$_B$ }}

\def\bj{\hbox{$b_j$ }}

\def\magsqsec{\hbox{mag~arcsec$^{-2}$ }}
\def\deg{\hbox{$^\circ$ }}
\def\x{\hbox{$\times$ }}
\def\4{\hbox{$4^\circ\times3^\circ$ }}
\def\-{\hbox{$\sim$}}
\def\<{\hbox{$<$ }}
\def\>{\hbox{$>$ }}
\title{The Surface Brightness and Colour-Magnitude Relations for Fornax Cluster Galaxies}
\author[A. M. Karick et al.]
       {A. M. Karick$^1$, M. J. Drinkwater$^2$ and M. D. Gregg$^{3,4}$\\
        $^1$School of Physics, University of Melbourne, Victoria 3010, Australia\\
        $^2$Department of Physics, University of Queensland, Queensland 4027, Australia\\
        $^3$Department of Physics, University of California, Davis, CA 95616, USA\\
	$^4$Institute for Geophysics and Planetary Physics, Lawrence Livermore National Laboratory, L-413, Livermore, CA 94550, USA\\}
\date{Draft version 1.0, July 14th 2002}

\pagerange{\pageref{firstpage}--\pageref{lastpage}}
\pubyear{2002}
\begin{document}
\maketitle
\label{firstpage}

\begin{abstract}
We present BVI photometry of 190 galaxies in the central \4 region of
the Fornax Cluster observed with the Michigan Curtis Schmidt
Telescope. Results from the Fornax Cluster Spectroscopic Survey (FCSS)
and the Flair-II Fornax Surveys have been used to confirm the
membership status of galaxies in the Fornax Cluster catalogue, FCC
\citep{Ferguson1989}.  In our catalogue of 213 member galaxies, 92
(43\%) have confirmed radial velocities.

In this paper we investigate the surface brightness-magnitude relation
for Fornax Cluster galaxies. Particular attention is given to the
sample of cluster dwarfs and the newly discovered ultra-compact dwarf
galaxies (UCDs) from the FCSS \citep{Drink2000,Deady2002}. We examine
the reliability of the surface brightness-magnitude relation as a
method for determining cluster membership and find that at surface
brightnesses fainter than 22 \magsqsec, it fails in its ability to
distinguish between cluster members and hardly resolved background
galaxies. Cluster members exhibit a strong surface
brightness-magnitude relation. Both elliptical (E) galaxies and dwarf
elliptical (dE) galaxies increase in surface brightness as luminosity
decreases. The UCDs lie off the locus of the relation.

B-V and V-I colours are determined for a sample of 113 cluster
galaxies and the colour-magnitude relation is explored for each
morphological type. The UCDs lie off the locus of the colour-magnitude
relation. Their mean V-I colours ($\sim$1.09) are similar to those of
globular clusters associated with NGC 1399. The location of the UCDs
on both surface brightness and colour-magnitude plots supports the
``galaxy threshing'' model for infalling nucleated dwarf elliptical (dE,N) galaxies \citep{Bekki2001}.
\end{abstract}


\section{Introduction}
In recent years the relationship between 'dwarf' galaxies and 'normal'
elliptical galaxies has been described as a 'dichotomy' - a
description based on the observation that each population exhibits a
unique surface brightness-luminosity relation \citep{Ferguson1994}. In
addition, surface brightness profiles differ between both
populations. Typical profiles for elliptical (E) galaxies follow a
$r^\frac{1}{4}$ power law whereas dwarf elliptical (dE) galaxies
generally follow an exponential law. Despite a lack of physical
understanding, the surface brightness-magnitude relation provides one
of the most fundamental means of galaxy classification and structure
determination.

Over the last few decades studies of galaxy clusters have been
dominated by wide-field photographic imaging. Population studies of
the Fornax Cluster
\citep{Bothun1991,Phillipps1987,Ferguson1988,Ferguson1989} have
traditionally combined image morphology and surface brightness
measurements to provide a statistical treatment of cluster membership.
By virtue of their low surface brightnesses, dwarf galaxies were
traditionally classified as cluster members whereas faint galaxies
with high surface brightness were assigned a background status.

Although studies of this nature have led to our present understanding
of cluster populations it seems that these traditional classification
techniques are not as robust as initially thought. Without known
radial velocities, compact elliptical galaxies such as M32, are
virtually indistinguishable from intrinsically luminous background E
galaxies \citep{Sandage1984}. The discovery of the large, low surface
brightness spiral Malin 1 \citep{Bothun1987} exemplifies the problem
of membership assignment on the basis of surface brightness. This
large background giant was previously classified as a foreground
cluster dwarf.

Simulations of galaxy evolution based on Cold-Dark-Matter (CDM) models
predict numbers of dwarf galaxies that far exceed those observed
locally \citep{Ferguson1994}. These models suggest that many of these low
surface brightness galaxies may be missing from magnitude-limited
surveys. Consequently our current understanding of the cluster
luminosity function may be significantly biased towards the 
high-luminosity limit. 
 
The nearby Fornax Cluster is one of the most well studied galaxy
clusters. Its environment and proximity provides a rich playing field
for galaxy studies over a large range of morphological types. Covering
an area of nearly 40 deg$^2$, Ferguson's optical catalogue of the
Fornax Cluster (FCC) \citep{Ferguson1989} contains 2678 galaxies of
which 340 are classified as ``likely cluster members''. Members were
originally identified visually from large-scale photographic plates,
primarily on the basis of morphology and surface brightness. Of the
340 ``likely members'', 186 were classified as dwarf galaxies and only
68 galaxies had known radial velocities.

Recently a spectroscopic survey of the Fornax Cluster
\citep{Drink2000:fss} revealed a small sample of ultra-compact dwarf
galaxies (UCDs) with radial velocities indicating cluster membership. In
previous studies using photographic plates, these unresolved high
luminosity objects were mis-classified as foreground stars.

In this paper we present BVI photometry of 190 Fornax Cluster members
obtained with the CTIO Curtis Schmidt Telescope.  After constructing a
\4 mosaic of the cluster we have measured the magnitudes and
surface-brightnesses of all cluster galaxies within the field. As a
result we have produced a surface brightness-magnitude relation that
includes a much larger range of galaxy luminosities and morphological
types than has previously been achieved. Since we have a large number of
galaxies with confirmed redshifts we have also investigated the use of
the surface brightness-magnitude relation for cluster membership
classification. Of particular interest is the high luminosity, high
surface brightness extension of the dwarf population relative to the
normal ellipticals and the location of the recently discovered
ultra-compact dwarf galaxies.

From the surface brightness-magnitude relation the relative positions
of the ultra-compact dwarfs to the remaining cluster population has
enabled us to further explore their origin. Our results support
analysis by \cite{Phillipps2001,Drink1999,Hilker2001} and others, that
the UCDs are a unique type of dwarf galaxy, possibly remnant nuclei of
infalling dwarfs which have been tidally stripped through their
interaction with the central cD galaxy, NGC 1399 \citep{Bekki2001}.

We have also determined the colours (B-V \& V-I) of the cluster
galaxies to investigate correlations with other photometric parameters
and identify a possible environmental influence on cluster galaxy
evolution.

In Section 2 we introduce our revised catalogue of Fornax Cluster
members in the central \4 region. Section 3 details our observations
and Section 4 the photometry of cluster members. We then discuss the
surface brightness-magnitude relations for cluster members and
non-members in Section 5 and galaxy colours in Section 6. We conclude
with a discussion and summary of our results in Sections 7 and 8.

\section{Cluster Galaxies}
Our sample of Fornax Cluster galaxies is largely based on Ferguson's
optical {\it Fornax Cluster Catalogue (FCC)} of 340 members. The 205
member galaxies from the FCC which lie in the central \4 region of the
cluster form the basis of our sample. In the original FCC only 52
(25\%) galaxies in the mosaic region were confirmed by their radial
velocities as cluster members. We have incorporated the results from a
number of recent spectroscopic surveys of the cluster
\citep{Drink1999,Hilker1999:II,Drink2001}, in which membership
classifications have been reassigned. Our final catalogue for the
central \4 region contains 213 cluster galaxies of which 92 (43\%)
have radial velocities.

\subsection{New radial velocities}
The development of the Two degree Field spectrograph (2dF) on the
Anglo-Australian Telescope has revolutionised cluster science. Unlike
photometric surveys, spectroscopy eliminates subjective judgements of
likely cluster membership based on morphology, surface brightness and
colour. The purpose of the {\it Fornax Cluster
Spectroscopic Survey, FCSS} (see Drinkwater et al. 1999) was to
confirm cluster membership (to a limiting magnitude of $b_j$ = 19.8)
from spectral analysis of {\it all} 14,000 objects (both resolved and
unresolved) in a 12 deg$^2$ region centred on NGC 1399, the optical
centre of the cluster. Recent results from the first $2^\circ$ diameter field
are presented in \cite{Deady2002}.

We have also incorporated the results from a spectroscopic survey by
\cite{Drink2001} of 675 bright (16.5 $<$ $b_j$ $<$ 18) galaxies in the
central 6\deg \x 6\deg region of the Fornax Cluster. Using the
FLAIR-II spectrograph on the UKST at the Anglo-Australian Observatory,
the purpose of this survey was to identify new cluster members which
may have been misidentified `by eye' as background objects, based on
their original surface brightness classifications. 

Table \ref{tab_newmembers} gives details of the FCC galaxies with
changed membership status resulting from both the FCSS and FLAIR-II
surveys. Morphological classifications are those adopted from
\cite{Drink2001} where the original FCC morphologies are converted to
``t-types'' following the definition of de Vaucouleurs et al
(1991).

\begin{table}
\caption{Galaxies listed in the FCC with changed membership classification.}
\begin{tabular}{lllll}
\hline
RA  \quad \quad \quad       Dec (J2000)        &  FCC      &   type  &   bj    &
 cz (km$s^{-1}$)\\
\hline                                
03:27:33.80 -35:43:04.0  &  470B     &  S/Im   &  17.5   & 723                \\
03:31:32.50 -35:03:43.0  &  729B     &  (d)SO  &  16.5   & 1676       \\
03:33:43.40 -35:51:33.0  &  123      &  ImV     &  17.9   & 15483             \\
03:33:56.20 -34:33:43.0  &  904B     &  (d)E   &  17.4   & 2254       \\
03:33:57.20 -34:36:43.0  &  905B     &  ?       &  17.7   & 1242              \\
03:34:57.27 -35:12:23.5  &  141      &  dE4     &  19.2   & 16735 $^1$ \\
03:36:42.83 -35:26:07.7  &  175      &  dE3     &  17.9   & 31430 $^1$ \\
03:37:08.27 -34:43:52.2  &  189      &  dE4,N   &  19.1   & 31044 $^1$ \\
03:38:16.75 -35:30:27.3  &  1241B    &  dE3?   &  14.7   & 2012 $^1$            \\
03:41:03.99 -35:38:51.2  &  257      &  dE0,N?  &  18.4  & 50391 $^1$ \\
03:41:59.50 -33:20:53.0  &  1554B    &  E       &  17.7   & 1642              \\
\hline
\label{tab_newmembers}
\end{tabular}\\
\small $^1$ Radial velocity taken from the FCSS.
\citep{Drink2000:fss}. \\All other velocity measurements are taken
from FLAIR-II survey \citep{Drink2001}. The morphological type of each
galaxy is taken from the original FCC \citep{Ferguson1989}.
\end{table}

Galaxies for which membership status has changed have also had their
``t-type'' morphological classification reassigned when
appropriate. For example the previously misidentified background
elliptical galaxy, FCCB1554 was confirmed by the Flair-II survey as a
cluster member and is therefore more likely to be a dwarf elliptical
galaxy with a -5 ``t-type'' classification.
 
\subsection{Discovery of the UCDs}
A small sample of ultra-compact dwarf galaxies (UCDs) were discovered
as part of the FCSS \citep {Drink1999,Deady2002}. Two of the objects
were previously identified by \cite{Hilker1999:I} in a study of the
globular clusters associated with NGC 1399. Assumed to be foreground
stars in earlier Fornax studies, spectral analysis of these very faint
objects gives radial velocities confirming them to be cluster members.
Optical imaging indicates they are unlike any other previously known
cluster dwarf, with intrinsic sizes of $\sim$ 100 pc and B band
magnitudes ranging from -13 to -11 \citep{Drink2000}.

The UCDs all lie within 30' of the central cD galaxy NGC 1399. The
most recent investigation of NGC 1399 globular clusters
\citep{Dirsch2003} indicates that their population extends out to more
than 25' at similar radii to the UCDs.

At present the origin of these enigmatic objects remains a
mystery. The two most favoured hypotheses for their origin are: (1)
globular cluster systems associated with the central cD galaxy NGC
1399, (2) the nuclei of tidally stripped infalling nucleated dwarf elliptical (dE,N) galaxies.
Observations support various aspects of both scenarios.
In either case their origin and evolution will provide a new insight
into the evolution of the faint cluster population.

\section{Data and observations}
Twelve 1.7 deg$^2$ CCD fields comprising the central $4^\circ$
$\times$ $3^\circ$ region of the Fornax Cluster were taken in
photometric conditions with the CTIO Curtis Schmidt Telescope. Using a
SITE 2048 \x2048 24 micron CCD each 1.3$^\circ$ \x 1.3$^\circ$ field
has 2.32'' pixels $\equiv$ 200 pc for d=20 Mpc
\citep{Drink2001ApJ}. Table \ref{tab_dates} details the observations
of each Fornax field.

\begin{table}
\caption{Details of the observations}
\begin{tabular}{llllll}
\hline
Date        & Field 	& RA \quad Dec (J2000)	& Filter& Exp.time(s) \\
\hline
1995 Nov 15 &  NW & 03:34:48 $-$34:26:48	& B 	& 300 $\times$ 9 \\
            &     &    		      	& V 	& 300 $\times$ 6 \\
            &     &    		      	& I 	& 300 $\times$ 6 \\
1995 Nov 16 &  N  & 03:39:39 $-$34:27:05	& B 	& 540 $\times$ 5 \\
            &     &    		      	& V 	& 360 $\times$ 5 \\
            &     &    		      	& I 	& 300 $\times$ 6 \\
1995 Nov 16 &  NE & 03:44:30 $-$34:27:23	& B 	& 420 $\times$ 6 \\
            &     &    		      	& V 	& 360 $\times$ 5 \\
            &     &    		      	& I 	& 360 $\times$ 5 \\
1995 Nov 17 &  E  & 03:42:38 $-$35:36:45	& B 	& 420 $\times$ 6 \\
            &     &    		      	& V 	& 360 $\times$ 5 \\
            &     &    		      	& I 	& 360 $\times$ 5 \\
1995 Nov 17 &  SE & 03:42:41 $-$36:36:45	& B 	& 420 $\times$ 6 \\
            &     &    		      	& V 	& 360 $\times$ 5 \\
            &     &    		      	& I 	& 360 $\times$ 5 \\
1995 Nov 18 &  S  & 03:39:37 $-$36:27:05	& B 	& 420 $\times$ 6 \\
            &     &    		      	& V 	& 360 $\times$ 5 \\
            &     &    		      	& I 	& 360 $\times$ 5 \\
1995 Nov 18 &  SW & 03:34:38 $-$36:26:47	& B 	& 420 $\times$ 6 \\
            &     &    		      	& V 	& 360 $\times$ 5 \\
            &     &    		      	& I 	& 360 $\times$ 5 \\
1995 Nov 19 &  NWW& 03:27:59 $-$34:36:45	& B 	& 420 $\times$ 6 \\
            &     &    		      	& V 	& 360 $\times$ 5 \\
            &     &    		      	& I 	& 360 $\times$ 5 \\
1995 Nov 19 &  WW & 03:27:52 $-$35:36:45	& B 	& 420 $\times$ 6 \\
            &     &    		      	& V 	& 360 $\times$ 5 \\
            &     &    		      	& I 	& 360 $\times$ 5 \\
1995 Nov 20 &  SWW& 03:29:35 $-$36:26:19	& B 	& 420 $\times$ 6 \\
            &     &    		      	& V 	& 360 $\times$ 5 \\
            &     &    		      	& I 	& 360 $\times$ 5 \\
1997 Jan 28 &  W  & 03:34:43 $-$35:26:48	& B 	& 420 $\times$ 6 \\
            &     &    		      	& V 	& 360 $\times$ 5 \\
            &     &    		      	& I 	& 360 $\times$ 5 \\
1997 Jan 29 & CORE& 03:39:38 $-$35:27:05	& B  	& 420 $\times$ 6 \\
            &           &    		      	& V     & 360 $\times$ 2 \\
            &           &    		      	& I	& 360 $\times$ 5 \\
\hline
\label{tab_dates}
\end{tabular}
\end{table}

Image reduction was carried out using the standard IRAF routines.
Photometric calibration was achieved using the Landolt UBVRI standard
star catalogue \citep{Landolt1992}. Typically two Landolt star fields
from a total of four (SA92, SA98, SA95 \& SA114) were observed each
night, each field containing 10-20 stars which were used for
calibration. The photometric data are therefore on the Cousins system.

Calibration of the individual fields was carried out using the IRAF
aperture photometry routines APPHOT and PHOTCAL. APPHOT was used to
measure the stellar aperture magnitudes (4'' radii) and PHOTCAL was
used to convert to calibrated magnitudes, correcting the instrumental
magnitudes for both airmass and colour. We found the colour terms for
this system were not significant compared to the typical uncertainties
of galaxy photometry ($\sim$ 0.1-0.2 mag) with this data.

Sky coordinates were fitted to the individual fields for all
filters. The IRAF routine CCMAP and the USNO-A V2.0 catalogue of
astrometric standards were used to compute the plate solution for each
image by pixel matching to celestial coordinates. The USNO-A V2.0
catalogue is favoured for its resulting accuracy in astrometry (mostly
in the reduction of systematic errors) and improved photometry (the
brightest stars on each plate have B and V magnitudes measured by the
Tycho experiment on the Hipparcos satellite). Typical residuals for the astrometry
were \-0.25''.

\begin{figure*}
\begin{center}
\includegraphics[scale=1.8]{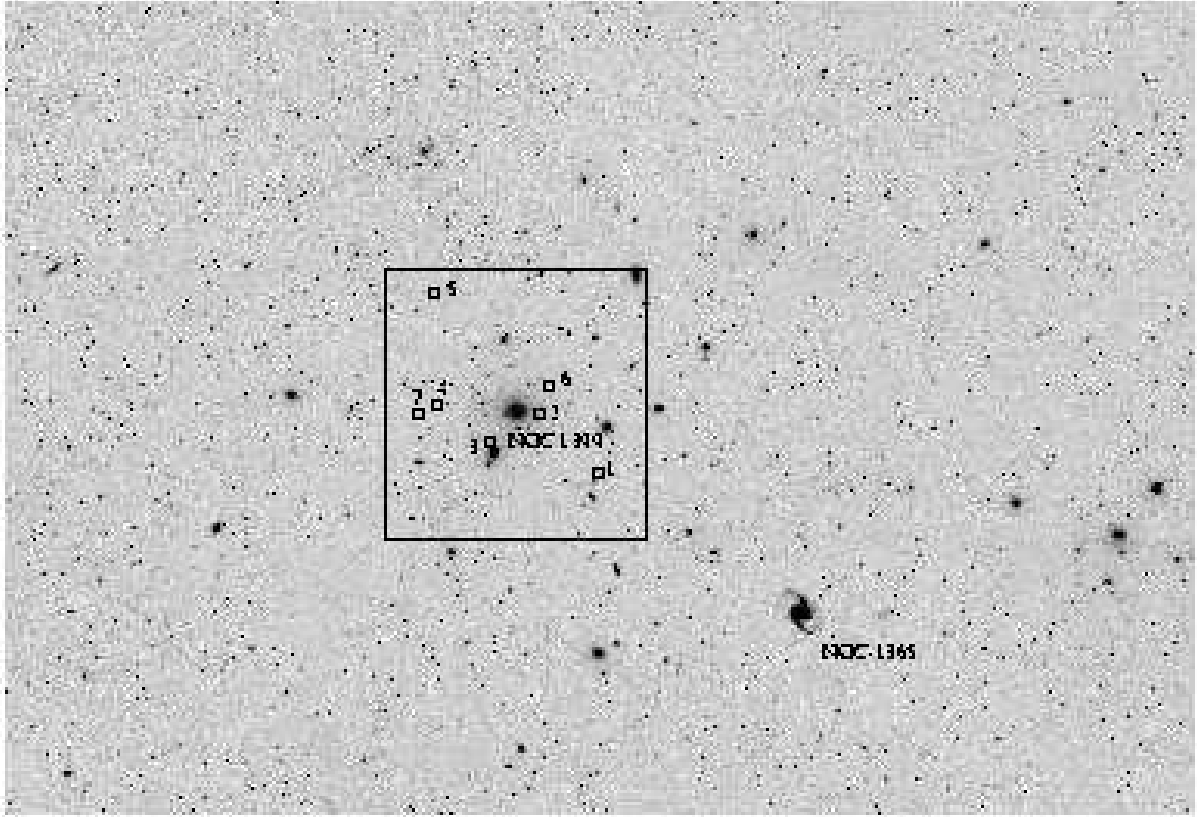}
\caption{B-band \4 Fornax Cluster mosaic with the distribution of the UCDs (boxes) in the central $1^\circ$ region. NGC 1399 marks the optical centre of the cluster. }
\label{fig_mosaic}
\end{center}
\end{figure*}

The images were then combined to form three $4^\circ$ $\times$
$3^\circ$ mosaics of the cluster in each filter (BVI). Figure
\ref{fig_mosaic} shows the B-band Fornax mosaic. The positions of the
UCDs are also shown. For each mosaic the sky coordinates of the
`Fornax Core' image were used as a reference coordinate system for the
entire cluster.

\section{Photometry of Cluster Members}
SExtractor (Source Extractor), is an automated program that optimally
detects, classifies and performs photometry on sources from
astronomical images \citep{Bertin1996}. Using the SExtractor package
we have compiled the photometry of the known Fornax Cluster members 
in our survey area. Their positions, magnitudes, peak
surface brightness and colours were measured using SExtractor
photometry routines.

Source Extraction is carried out in five main steps. An estimation of
the sky background is made before a detection threshold
is set. A median filter of 3 \x 3 pixels was applied in order to
suppress overestimations due to foreground stars. The images were then
convolved with a Gaussian of 1.5'' FWHM. For each mosaic image an
object detection threshold was set to 2$\sigma$ above the local sky
background level in order to detect the faintest objects in the catalogue (\bj
\- 19.8, 24.9 B \magsqsec). A minimum number of 6 connected pixels was also required for
a positive detection.

\begin{figure}
\begin{center}
\includegraphics{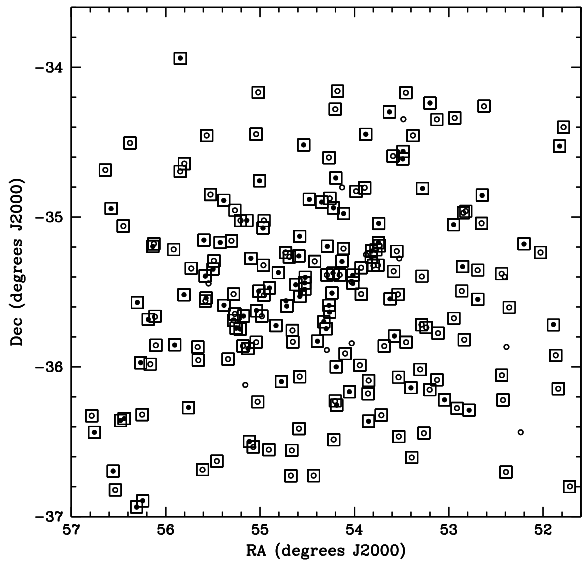}
\caption{Distribution of cluster members in the \4 survey area;
galaxies with radial velocities (filled circles), unconfirmed members (open circles).
Boxes indicate SExtractor galaxy detections in all filters (BVI).}
\label{fig_detections}
\end{center}
\end{figure}

Filtering of the \- 60,000 SExtractor detections was achieved by
matching the SExtractor catalogue to the Fornax Cluster members in the
revised catalogue. All detections were confirmed by eye with obvious
mis-identifications removed. Figure \ref{fig_detections} shows the member
galaxies detected from our revised catalogue. 

Total magnitudes were calculated using the SExtractor MAG\_BEST
option, an aperture photometry routine inspired by Kron's ``first
moment'' algorithm. This routine corrects for overcrowding to which
aperture photometry is sensitive \citep{Bertin1996}. For isolated
objects aperture magnitudes are calculated however in crowded areas
where objects may overlap, an isophotal magnitude is calculated where
the fraction of flux lost by the isophotal magnitude is estimated and
corrected. This assumes the intensity profiles have Gaussian wings
resulting from atmospheric blurring \citep{Hilker1999:I}. 

We have used a fairly simple method for determining the central
surface brightness of galaxies and have made no attempt to fit model
profiles. This is partly due to the low resolution of the data as well
as our attempt to reproduce the surface-brightness relation without
making assumptions about galaxy morphology. Our peak surface
brightness is simply calculated by determining the peak flux in the
central pixel and dividing by the pixel area. As a result the surface
brightness of a galaxy will be largely influenced by the seeing and
our measured peak surface brightness will be an underestimate of the
true central surface brightness. The measured peak surface brightness
will be consistent with the true central surface brightness only for
galaxies with a slow varying surface brightness at their centres, such
as exponential profile galaxies with a large scale length. The
implications of this will be discussed later.

B-V and V-I colours were determined using aperture photometry with
radii = 3.5$r_{kron}$ which contains 99.3\% of the light for galaxies
with exponential profiles. Apertures defined by running the source
detection algorithm on the V-band image, were then used for the B \&
I-band photometry. This ensures that galaxy colours are determined
using identical apertures in all three filters. Table \ref{tab_resa}
contains the photometric properties of the Fornax members detected.

Photometry of the UCDs was done using the IRAF aperture photometry
routines since the majority of them fell below the SExtractor
detection threshold. Magnitudes and surface brightnesses were
determined using the same techniques as our SExtractor photometry. The
photometry of the UCDs is given in Table \ref{tab_ucds}. Note that
UCD2 may be suffering contamination by a foreground star. The central
surface brightness is also 2 mag arcsec$^{-2}$ brighter than those of the
remaining UCD population.  The colours for UCD5 are difficult to
interpret. The errors on the measurements are quite large and we find a
very blue V-I value given its B-V colour. The V-I colour for UCD5 is
significantly bluer than that of the remaining UCDs, which have been independently
measured by \cite{Mieske2002} ($\langle$V-I$\rangle$=1.13).  This is
an unexpected result and we suspect that higher resolution
observations will yield a more accurate colour measurement of this enigmatic object.
The accuracy of the photometry is limited by the large pixels (2.32'')
which makes photometry, particularly of the smallest dwarfs
difficult. Our photometric errors in all bands are typically \-0.1-0.2
mag.
 
\begin{table*}
\caption{Photometry of the ultra-compact dwarf galaxies discovered in the FCSS}
\begin{tabular}{lllllllll}
\hline
FCSS Name  (UCD\#) & RA \quad \quad Dec (J2000) &  cz &         $V$ & $B-V$ & $V-I$ & $\mu_{B peak}$ & $\mu_{V peak}$ & $\mu_{I peak}$ \\
                  &                         &   (km$s^{-1}$) & & & \\
\hline
J033703.3$-$353804 (1) &03:37:03.30  -35:38:04.6  & 1481      & 19.3   & 0.9    & 1.2   & 24.9    & 22.2    &  22.1   \\     
J033806.3$-$352858 (2) &03:38:06.33  -35:28:58.8  & 1312      & 19.0   & 0.9    & 1.0   & 21.1    & 19.4    &  19.8   \\     
J033854.1$-$353333 (3) &03:38:54.10  -35:33:33.6  & 1468 $^1$ & 17.7   & 1.1    & 1.3   & 23.2    & 21.0    &  20.5$^3$\\
J033935.9$-$352824 (4) &03:39:35.95  -35:28:24.5  & 1920 $^2$ & 18.8   & 0.8    & 1.0   & 24.3    & 22.2    &  21.9   \\     
J033952.5$-$350424 (5) &03:39:52.58  -35:04:24.1  & 1355      & 19.0   & 1.0    & 0.8   & 25.1    & 22.6    &  22.5$^3$\\
J033805.0$-$352409 (6) &03:38:05.08  -35:24:09.6  & 1211      & 18.9   & 0.8    & 1.2   & 23.9    & 21.8    &  21.1   \\     
J034003.4$-$352944 (7) &03:40:03.40  -35:29:44.0  & 2251      & 18.4   & 0.9    & 1.1   & 23.5    & 21.0    &  21.2   \\     
\hline
\label{tab_ucds}
\end{tabular}\\
\small  $^1$ CGF 1-4 and $^2$ CGF 5-4, in \cite{Hilker1999:I}. $^3$
Measurements taken using 6'' radii apertures.\\ All other measurements
taken with 4'' radii apertures. Colours are in the Cousins system.
\end{table*}

\begin{table*}
\caption{Photometry of Fornax Cluster members}
{\tiny
\begin{tabular}{lllllllllll}
\hline
FCC &RA \quad \quad Dec (J2000)&$V_{best}$ & $\mu_{V peak}$ & $B_{best}$ & $\mu_{B peak}$ & $I_{best}$ &$\mu_{I peak}$& $B-V$ $^1$ & $V-I$ $^1$& ``t-type''\\
\hline
\hline
52   &	3:27:08.14	-34:23:59.3	& 18.17	& 23.88	& 18.89	& 24.74	& 16.60	& 22.95	& 0.68 	& 1.09 	& -5\\
55   &	3:27:18.05	-34:31:33.9	& 12.91	& 19.38	& 14.04	& 20.34	& 11.46	& 18.37	& 0.93 	& 1.43 	& -1\\
56   &	3:27:21.61	-36:08:47.2	& 17.08	& 23.25	& 17.81	& 23.86	& 15.92	& 22.28	& --	& --	& -5\\
57   &	3:27:27.91	-35:55:24.5	& 17.16	& 22.90	& 17.79	& 23.48	& 16.26	& 22.04	& 0.51 	& 0.99 	& -5\\
65   &	3:28:06.68	-35:14:11.7	& 16.95	& 22.79	& 17.75	& 23.48	& 16.01	& 21.86	& 0.66 	& 1.00 	& -5\\
67   &	3:28:48.77	-35:10:49.1	& 13.37	& 20.93	& 14.15	& 21.93	& 12.13	& 19.62	& 0.79 	& 1.24 	& 5 \\
71   &	3:29:26.30	-35:36:13.3	& 19.08	& 24.72	& --	& --	& --	& --	& --	& --	& -5\\
73   &	3:29:34.53	-36:42:10.8	& 19.12	& 24.52	& 20.09	& 24.75	& 18.22	& 23.80	& --	& --	& -5\\
77   &	3:29:42.00	-36:13:17.2	& 17.98	& 23.42	& 18.55	& 24.68	& 17.07	& 23.05	& 0.69 	& 1.02 	& -5\\
78   &	3:29:45.47	-35:22:42.4	& 18.56	& 24.48	& --	& --	& --	& --	& 0.61 	& 1.18 	& -5\\
82   &	3:30:30.52	-34:15:36.0	& 15.87	& 22.17	& 16.69	& 22.88	& 14.79	& 21.10	& 0.77 	& 1.10 	& -5\\
83   &	3:30:35.19	-34:51:19.2	& 11.35	& 17.13	& 12.30	& 18.56	& 10.11	& 16.16	& 0.92 	& 1.24 	& -4\\
84   &	3:30:36.64	-35:02:29.3	& 18.76	& 24.37	& 19.51	& 24.59	& 17.19	& 22.68	& --	& --	& -5\\
85   &	3:30:46.21	-35:32:57.9	& 15.65	& 22.07	& 16.43	& 22.98	& 14.61	& 21.12	& 0.81 	& 1.13 	& -5\\
86   &	3:30:46.68	-35:21:15.9	& 17.12	& 23.18	& --	& --	& --	& --	& 0.76 	& 1.18 	& -5\\
90   &	3:31:08.21	-36:17:24.2	& 14.19	& 19.33	& 14.82	& 19.84	& 13.16	& 18.81	& --	& --	& -4\\
92   &	3:31:16.86	-34:57:37.7	& 18.66	& 24.51	& --	& --	& --	& --	& --	& --	& 10\\
93   &	3:31:20.75	-35:49:09.9	& 18.97	& 24.69	& --	& --	& 17.87	& 23.50	& --	& --	& -5\\
94   &	3:31:22.84	-34:58:18.0	& 19.03	& 24.68	& 19.85	& 25.10	& 17.18	& 23.53	& --	& --	& -5\\
95   &	3:31:24.83	-35:19:51.1	& 14.02	& 20.14	& 14.90	& 21.11	& 12.85	& 19.10	& 0.85 	& 1.18 	& -1\\
97   &	3:31:26.90	-35:29:40.3	& 18.65	& 24.55	& --	& --	& --	& --	& --	& --	& -5\\
98   &	3:31:39.17	-36:16:35.0	& 18.41	& 24.49	& --	& --	& 18.22	& 23.55	& --	& --	& 10\\
99   &	3:31:44.96	-34:20:17.3	& 16.58	& 23.19	& 17.41	& 23.85	& 15.73	& 22.25	& 0.72 	& 0.97 	& -5\\
101  &	3:31:46.90	-35:40:32.4	& 16.56	& 22.76	& 17.17	& 23.72	& --	& --	& --	& --	& -5\\
100  &	3:31:47.67	-35:03:05.5	& 14.86	& 21.84	& 15.75	& 22.65	& 13.72	& 20.79	& 0.85 	& 1.13 	& -5\\
102  &	3:32:10.78	-36:13:13.8	& 15.97	& 22.49	& 16.42	& 22.88	& 15.00	& 21.86	& --	& --	& 10\\
103  &	3:32:27.38	-35:46:30.8	& 19.80	& 24.35	& --	& --	& 18.35	& 23.76	& --	& --	& -5\\
105  &	3:32:29.93	-36:05:17.0	& 17.60	& 23.77	& 17.13	& 24.43	& 16.19	& 22.78	& --	& --	& -5\\
104  &	3:32:30.36	-34:20:52.9	& --	& --	& 19.16	& 24.43	& --	& --	& --	& --	& -5\\
106  &	3:32:47.77	-34:14:18.9	& 14.52	& 20.07	& 15.49	& 21.16	& 13.40	& 19.05	& 0.97 	& 1.12 	& -1\\
108  &	3:32:48.69	-36:09:11.0	& 19.12	& 24.54	& 18.86	& 24.90	& 18.12	& 23.48	& --	& --	& -5\\
110  &	3:32:57.37	-35:44:15.5	& 16.69	& 23.17	& 17.39	& 24.15	& 15.25	& 22.23	& 0.75 	& 1.23 	& -5\\
112  &	3:33:03.34	-36:26:35.9	& 16.67	& 23.12	& 17.29	& 23.83	& 15.73	& 22.25	& 0.64 	& 1.02 	& -5\\
113  &	3:33:06.83	-34:48:31.7	& 14.83	& 21.71	& 15.51	& 22.42	& 13.84	& 20.87	& 0.66 	& 0.98 	& 5 \\
115  &	3:33:09.27	-35:43:05.2	& 16.31	& 22.79	& 16.71	& 23.17	& 15.34	& 22.17	& 0.41 	& 0.91 	& 8 \\
116  &	3:33:12.83	-36:01:02.4	& 15.77	& 22.06	& 16.46	& 22.77	& 14.69	& 21.13	& 0.80 	& 1.05 	& -5\\
118  &	3:33:31.29	-34:27:19.5	& 17.28	& 23.40	& 18.14	& 24.15	& 16.45	& 22.54	& --	& --	& -5\\
120  &	3:33:34.24	-36:36:19.7	& 15.99	& 22.74	& 16.46	& 23.04	& 15.04	& 22.04	& 0.44 	& 0.95 	& 10\\
121  &	3:33:36.31	-36:08:27.8	&  9.31	& 17.54	& 10.23	& 18.55	&  8.18	& 16.13	& 0.65 	& 1.14 	& 4 \\
125  &	3:33:48.52	-35:50:09.5	& 18.95	& 24.17	& 19.82	& 24.89	& 17.28	& 23.53	& --	& --	& -5\\
124  &	3:33:49.04	-34:10:10.5	& 17.85	& 23.92	& 18.80	& 24.40	& 16.97	& 23.04	& --	& --	& -5\\
904  &	3:33:56.20	-34:33:43,0	& 16.66	& 21.86	& 17.57	& 22.75	& 15.60	& 21.01	& 0.89 	& 1.05 	& -5\\
905  &	3:33:57.20	-34:36:40.0	& 16.83	& 22.41	& 17.53	& 23.22	& 15.93	& 21.87	& 0.73 	& 0.96 	& 10\\
128  &	3:34:07.07	-36:27:56.6	& 16.35	& 22.42	& 16.94	& 22.89	& 15.50	& 21.69	& 0.54 	& 1.00 	& 10\\
129  &	3:34:07.74	-36:04:10.7	& 19.21	& 24.51	& --	& --	& --	& --	& --	& --	& 10\\
130  &	3:34:09.23	-35:30:59.7	& 18.99	& 24.58	& --	& --	& 16.70	& 23.52	& --	& --	& 10\\
131  &	3:34:12.23	-35:13:40.9	& 20.30	& 24.42	& --	& --	& 18.92	& 23.66	& --	& --	& -5\\
132  &	3:34:18.31	-35:47:40.3	& 17.91	& 23.12	& 18.78	& 23.85	& 16.86	& 22.18	& 0.87 	& 0.99 	& -5\\
133  &	3:34:20.22	-35:21:43.4	& --	& --	& 17.70	& 24.10	& --	& --	& --	& --	& -5\\
134  &	3:34:21.80	-34:35:33.4	& 17.09	& 22.58	& 17.97	& 23.35	& 15.96	& 21.75	& --	& --	& -5\\
136  &	3:34:29.54	-35:32:45.9	& 14.17	& 20.55	& 15.02	& 21.58	& 13.04	& 19.58	& 0.82 	& 1.13 	& -5\\
135  &	3:34:30.89	-34:17:51.0	& 15.11	& 21.26	& 16.00	& 22.32	& 14.00	& 20.28	& 0.89 	& 1.10 	& -5\\
137  &	3:34:44.16	-35:51:40.8	& 17.26	& 23.64	& 17.40	& 24.61	& 16.42	& 22.80	& 0.39 	& 0.99 	& -5\\
140  &	3:34:56.49	-35:11:27.5	& 18.43	& 24.08	& 19.20	& 24.85	& 16.96	& 23.25	& 0.66 	& 1.05 	& -5\\
142  &	3:34:58.30	-35:02:32.6	& 18.21	& 24.06	& 19.50	& 25.08	& 17.01	& 23.17	& --	& --	& -5\\
143  &	3:34:59.21	-35:10:14.6	& 13.38	& 18.51	& 14.28	& 19.65	& 12.19	& 17.56	& 0.89 	& 1.19 	& -4\\
144  &	3:35:00.30	-35:19:19.7	& 18.82	& 24.06	& 19.82	& 24.75	& 17.82	& 23.45	& --	& --	& -5\\
145  &	3:35:05.54	-35:13:06.0	& --	& --	& 18.77	& 24.76	& --	& --	& --	& --	& -5\\
146  &	3:35:11.58	-35:19:22.4	& 19.48	& 24.24	& 20.37	& 24.89	& 18.67	& 23.58	& --	& --	& -5\\
147  &	3:35:16.90	-35:13:38.7	& 11.07	& 17.19	& 12.03	& 18.37	& 9.82	& 16.14	& 0.93 	& 1.25 	& -4\\
148  &	3:35:16.94	-35:16:00.7	& 12.32	& 18.15	& 13.22	& 19.05	& 11.18	& 17.30	& 0.82 	& 1.15 	& -1\\
149  &	3:35:23.85	-36:05:29.1	& 20.07	& 24.97	& --	& --	& 14.00	& 19.59	& --	& --	& -5\\
150  &	3:35:24.06	-36:21:49.2	& 15.13	& 20.55	& 15.86	& 21.44	& 19.16	& 23.49	& 0.74 	& 1.12 	& -5\\
151  &	3:35:25.42	-36:10:43.2	& 17.48	& 22.75	& 17.95	& 23.90	& 16.27	& 21.26	& --	& --	& -5\\
153  &	3:35:31.06	-34:26:49.5	& 12.27	& 18.39	& 13.35	& 19.66	& 11.04	& 17.38	& 1.07 	& 1.22 	& -1\\
155  &	3:35:33.96	-34:48:16.7	& 18.21	& 23.64	& 19.06	& 24.60	& 17.22	& 22.81	& --	& --	& -5\\
156  &	3:35:42.79	-35:20:17.2	& 17.99	& 23.34	& 19.32	& 24.37	& 16.85	& 22.37	& --	& --	& -5\\
157  &	3:35:42.84	-35:30:50.2	& 16.89	& 23.43	& 17.72	& 24.27	& 15.82	& 22.52	& 0.71 	& 1.20 	& -5\\
158  &	3:35:46.35	-35:59:22.4	& 16.27	& 23.00	& 17.02	& 23.86	& 15.40	& 22.10	& 0.71 	& 1.09 	& -5\\
159  &	3:35:55.68	-34:49:39.9	& 18.16	& 24.03	& 19.42	& 25.05	& 17.17	& 23.49	& --	& --	& -5\\
160  &	3:36:04.07	-35:23:19.5	& 16.89	& 22.01	& 17.72	& 23.59	& 15.78	& 20.38	& --	& --	& -5\\
161  &	3:36:04.10	-35:26:34.5	& 11.10	& 17.67	& 12.04	& 18.90	& 9.87	& 16.60	& 0.92 	& 1.23 	& -4\\
164  &	3:36:12.94	-36:09:59.0	& 15.83	& 21.50	& 16.50	& 22.33	& 14.74	& 20.56	& 0.71 	& 1.08 	& -5\\
165  &	3:36:23.59	-35:54:40.6	& 17.09	& 23.39	& 17.76	& 24.14	& 16.08	& 22.56	& --	& --	& -5\\
167  &	3:36:27.61	-34:58:35.8	& 10.05	& 17.05	& 11.09	& 18.30	& 8.78	& 16.02	& 1.03 	& 1.26 	& 0 \\
168  &	3:36:27.88	-35:12:37.9	& 18.68	& 24.14	& 19.20	& 24.80	& 17.23	& 22.95	& --	& --	& -5\\
170  &	3:36:31.75	-35:17:48.1	& 11.50	& 17.17	& 12.46	& 18.31	& 10.24	& 16.01	& 0.95 	& 1.25 	& -1\\
171  &	3:36:37.51	-35:23:09.4	& 18.15	& 23.83	& 19.34	& 24.47	& 16.85	& 23.17	& --	& --	& -5\\
173  &	3:36:43.12	-34:09:32.7	& 17.68	& 23.36	& 18.70	& 24.39	& 16.60	& 22.50	& --	& --	& -5\\
176  &	3:36:45.14	-36:15:21.9	& 12.15	& 19.24	& 12.32	& 20.44	& 11.37	& 18.52	& --	& --	& 1 \\
179  &	3:36:46.41	-36:00:02.0	& 11.16	& 17.16	& 12.02	& 18.34	& 9.94	& 16.16	& 0.86 	& 1.23 	& 1 \\
177  &	3:36:47.50	-34:44:21.0	& 12.57	& 18.96	& 13.63	& 20.12	& 11.37	& 18.06	& 1.04 	& 1.19 	& -1\\
178  &	3:36:48.64	-34:16:48.0	& 16.75	& 23.06	& 17.60	& 23.97	& 15.66	& 22.05	& 0.82 	& 1.07 	& -5\\
183  &	3:36:52.89	-36:29:10.4	& 16.76	& 22.91	& 17.36	& 23.73	& 15.60	& 22.02	& 0.71 	& 1.07 	& -5\\
181  &	3:36:53.21	-34:56:17.3	& 17.04	& 22.78	& 17.56	& 23.65	& 16.05	& 21.87	& 0.59 	& 1.00 	& -5\\
182  &	3:36:54.39	-35:22:27.4	& 14.07	& 19.53	& 14.95	& 20.64	& 12.91	& 18.70	& 0.88 	& 1.16 	& -1\\
184  &	3:36:57.00	-35:30:28.6	& 10.67	& 16.78	& 11.66	& 18.01	& 9.37	& 16.11	& 0.98 	& 1.29 	& -1\\
185  &	3:37:02.78	-34:52:30.9	& 19.44	& 24.37	& --	& --	& 18.18	& 23.52	& --	& --	& -5\\
188  &	3:37:04.66	-35:35:24.0	& 16.41	& 21.66	& 16.96	& 23.01	& 14.49	& 21.00	& 0.64 	& 0.94 	& -5\\
187  &	3:37:04.86	-34:36:11.0	& 15.62	& 21.93	& 16.35	& 22.81	& 15.59	& 21.03	& 0.75 	& 1.13 	& -5\\
190  &	3:37:09.02	-35:11:42.3	& 12.76	& 19.12	& 13.70	& 20.21	& 11.56	& 18.10	& 0.91 	& 1.20 	& -1\\
191  &	3:37:10.04	-35:23:12.3	& 19.25	& 24.28	& 20.17	& 25.18	& 18.02	& 23.35	& --	& --	& -5\\
193  &	3:37:11.83	-35:44:44.5	& 11.53	& 17.32	& 12.46	& 18.37	& 10.29	& 16.28	& 0.90 	& 1.22  & -1\\
194  &	3:37:17.98	-35:41:54.8	& 17.57	& 23.14	& 17.90	& 23.94	& --	& --	& 0.75 	& 1.09 	& -5\\
195  &	3:37:23.41	-34:54:00.1	& 16.35	& 22.73	& 17.11	& 23.49	& 15.18	& 21.68	& 0.77 	& 1.09 	& -5\\
\hline
\label{tab_resa}
\end{tabular}\\
\small $^1$ Colours are in the Cousins system.
}
\end{table*}

\begin{table*}
\caption{(cont'd) Photometry of Fornax Cluster members}
{\tiny
\begin{tabular}{lllllllllll}
\hline
FCC &RA \quad \quad Dec (J2000)&$V_{best}$ & $\mu_{V peak}$ & $B_{best}$ & $\mu_{B peak}$ & $I_{best}$ &$\mu_{I peak}$& $B-V$ $^1$ & $V-I$ $^1$& ``t-type''\\
\hline
\hline
196  &	3:37:33.96	-35:49:44.8	& 17.17	& 23.31	& 17.79	& 24.02	& 16.17	& 22.35	& 0.62 	& 1.18 	& -5\\
197  &	3:37:41.01	-35:17:46.2	& 18.91	& 23.95	& 19.71	& 25.18	& 18.51	& 23.44	& --	& --	& -5\\
199  &	3:37:43.73	-36:43:35.4	& 18.22	& 24.13	& 19.13	& 24.74	& 17.38	& 23.50	& 0.69 	& 0.97 	& 10\\
200  &	3:37:54.77	-34:52:55.0	& 16.85	& 22.64	& 17.37	& 23.15	& 15.71	& 21.95	& 0.49 	& 0.96 	& -5\\
202  &	3:38:06.55	-35:26:22.7	& 14.01	& 20.49	& 11.63	& 19.92	& --	& --	& 0.86 	& 1.19 	& -4\\
203  &	3:38:09.25	-34:31:05.8	& 14.78	& 21.09	& 15.28	& 21.78	& 13.86	& 20.31	& 0.50 	& 0.94 	& -5\\
208  &	3:38:18.80	-35:31:49.4	& 16.58	& 22.70	& 16.84	& 23.52	& 15.63	& 21.85	& 0.59 	& 0.98 	& -5\\
210  &	3:38:19.22	-36:03:56.5	& 18.79	& 23.95	& 19.14	& 24.72	& 14.22	& 20.14	& 0.53 	& 1.11 	& -5\\
207  &	3:38:19.27	-35:07:43.4	& 15.30	& 21.05	& 16.10	& 21.88	& 17.64	& 23.25	& 0.79 	& 1.06 	& -5\\
212  &	3:38:21.01	-36:24:47.6	& 16.93	& 23.48	& 18.05	& 24.28	& 16.36	& 22.74	& --	& --	& -5\\
211  &	3:38:21.48	-35:15:34.6	& 15.77	& 20.97	& 16.52	& 21.80	& 14.71	& 20.15	& 0.75 	& 1.06 	& -4\\
213  &	3:38:29.29	-35:27:07.0	& 9.32	& 16.59	& 10.77	& 17.69	& --	& --	& --	& --	& -4\\
214  &	3:38:36.62	-35:50:01.5	& 18.95	& 23.74	& 19.21	& 24.51	& 17.57	& 23.25	& 0.51 	& 0.10 	& -5\\
215  &	3:38:37.63	-35:45:24.6	& 19.45	& 24.35	& 19.97	& 24.91	& 18.60	& 23.53	& 0.61 	& 0.97 	& -5\\
216  &	3:38:39.26	-36:33:28.7	& 18.23	& 24.47	& 19.56	& 24.56	& 16.54	& 23.24	& --	& --	& -5\\
217  &	3:38:41.61	-36:43:36.8	& 19.50	& 24.49	& 20.38	& 24.90	& 18.25	& 23.37	& --	& --	& -5\\
218  &	3:38:45.42	-35:15:57.0	& 18.15	& 23.80	& 18.69	& 24.45	& 17.05	& 22.96	& 0.59 	& 1.17 	& -5\\
219  &	3:38:52.24	-35:35:42.4	& 9.92	& 16.39	& 10.88	& 17.48	& 8.66	& 15.91	& 0.97 	& 1.25 	& -4\\
220  &	3:38:55.15	-35:14:11.6	& 18.11	& 23.98	& 19.39	& 25.00	& 16.42	& 23.21	& --	& --	& -5\\
221  &	3:39:05.78	-36:05:55.2	& 17.13	& 22.44	& 17.76	& 23.21	& 16.15	& 21.53	& 0.63 	& 1.05 	& -5\\
222  &	3:39:13.42	-35:22:15.7	& 14.03	& 21.54	& 15.52	& 22.34	& 12.86	& 20.50	& 0.93 	& 1.30 	& -5\\
227  &	3:39:50.23	-35:31:20.9	& 19.40	& 24.42	& 19.85	& 24.61	& 17.35	& 23.55	& --	& --	& -5\\
226  &	3:39:50.24	-35:01:20.8	& --	& --	& --	& --	& 18.42	& 23.48	& --	& --	& -5\\
228  &	3:39:51.41	-35:19:18.9	& 18.96	& 24.12	& --	& --	& 18.35	& 23.57	& --	& --	& -5\\
229  &	3:39:55.31	-35:39:44.2	& 18.24	& 24.28	& 19.19	& 24.59	& 17.69	& 23.33	& --	& --	& -5\\
230  &	3:40:01.30	-34:45:28.5	& 16.80	& 22.35	& 17.14	& 22.89	& 15.84	& 21.54	& 0.39 	& 0.93 	& -5\\
231  &	3:40:04.62	-34:10:02.7	& 17.80	& 23.19	& 18.31	& 24.33	& 16.89	& 22.41	& --	& --	& -5\\
233  &	3:40:06.26	-36:14:00.8	& --	& --	& 19.89	& 24.83	& --	& --	& --	& --	& 10\\
236  &	3:40:09.93	-35:50:10.1	& 18.34	& 24.04	& 18.93	& 24.77	& --	& --	& --	& --	& -5\\
234  &	3:40:10.02	-34:26:46.0	& 16.60	& 23.29	& 17.12	& 23.78	& 15.80	& 22.57	& 0.48 	& 0.80 	& -5\\
238  &	3:40:17.19	-36:32:05.5	& 17.97	& 23.81	& 19.23	& 24.77	& 17.16	& 22.97	& --	& --	& -5\\
241  &	3:40:23.41	-35:16:32.8	& 16.22	& 22.89	& 16.96	& 23.70	& 15.30	& 21.91	& 0.75 	& 1.01 	& -5\\
243  &	3:40:27.02	-36:29:56.1	& 15.81	& 22.09	& 16.62	& 22.90	& 14.79	& 21.07	& 0.79 	& 1.05 	& -5\\
244  &	3:40:30.73	-35:52:39.3	& 17.69	& 23.16	& 18.44	& 23.85	& 16.36	& 21.42	& --	& --	& -5\\
245  &	3:40:33.86	-35:01:21.4	& 15.37	& 21.72	& 15.94	& 22.45	& 14.40	& 20.90	& 0.60 	& 1.05 	& -5\\
247  &	3:40:42.32	-35:39:40.0	& 17.67	& 23.63	& 18.00	& 23.75	& 17.08	& 23.00	& --	& --	& -5\\
248  &	3:40:43.43	-35:51:39.0	& 18.40	& 23.30	& 19.05	& 24.00	& 17.13	& 22.46	& 0.69 	& 1.14 	& -5\\
251  &	3:40:49.62	-35:01:24.4	& 18.79	& 24.37	& 18.86	& 24.95	& 17.86	& 23.52	& --	& --	& -5\\
252  &	3:40:50.38	-35:44:53.5	& 15.32	& 21.16	& 16.22	& 22.17	& 14.18	& 20.26	& 0.89 	& 1.152	& -5\\
254  &	3:41:00.77	-35:44:31.1	& 16.59	& 23.09	& 18.04	& 23.85	& 16.10	& 22.35	& --	& --	& -5\\
256  &	3:41:03.80	-34:57:16.2	& 19.23	& 24.60	& --	& --	& --	& --	& --	& --	& -5\\
258  &	3:41:07.13	-35:41:24.4	& 20.70	& 24.63	& --	& --	& 18.18	& 23.37	& --	& --	& -5\\
259  &	3:41:07.48	-35:30:53.5	& 17.58	& 23.69	& 18.51	& 24.65	& 17.24	& 22.77	& --	& --	& -5\\
260  &	3:41:12.79	-35:09:29.8	& 16.67	& 22.88	& 17.34	& 23.84	& 15.79	& 22.14	& 0.66 	& 0.86 	& -5\\
262  &	3:41:21.42	-35:56:54.3	& 17.44	& 23.64	& 18.04	& 24.24	& 16.01	& 21.47	& --	& --	& -5\\
264  &	3:41:31.83	-35:35:20.9	& 16.23	& 21.89	& 16.92	& 22.60	& 15.16	& 21.09	& 0.70 	& 1.02 	& -5\\
263  &	3:41:32.33	-34:53:21.9	& 13.39	& 19.75	& 13.65	& 19.93	& 12.60	& 19.18	& --	& --	& 6 \\
266  &	3:41:41.31	-35:10:12.5	& 15.72	& 20.91	& 16.02	& 21.90	& 14.16	& 19.91	& 0.79 	& 1.17 	& -5\\
268  &	3:41:49.86	-36:37:48.1	& 18.50	& 23.53	& 19.43	& 24.44	& --	& --	& --	& --	& -5\\
269  &	3:41:57.41	-35:17:31.4	& 18.17	& 24.06	& --	& --	& 17.24	& 22.64	& 0.70 	& 1.14 	& -5\\
1554  &	3:41:59.50	-35:20:53.0	& 16.23	& 21.89	& --	& --	& 14.98	& 20.85	& 0.90 	& 1.29 	& -5\\
271  &	3:42:06.23	-34:50:58.0	& 18.43	& 23.86	& 19.29	& 24.57	& 17.24	& 23.29	& 0.67 	& 1.01 	& -5\\
273  &	3:42:15.77	-34:27:17.5	& 18.04	& 23.84	& 19.15	& 24.20	& 17.48	& 22.88	& 0.60 	& 0.98 	& -5\\
274  &	3:42:17.31	-35:32:25.7	& 16.01	& 21.82	& 16.54	& 22.84	& 15.02	& 21.26	& 0.65 	& 0.99 	& -5\\
275  &	3:42:18.98	-35:33:38.7	& --	& --	& 18.56	& 24.79	& 17.64	& 23.38	& --	& --	& -5\\
276  &	3:42:19.32	-35:23:40.8	& 10.86	& 17.07	& 11.77	& 18.21	& 9.64	& 16.05	& 0.89 	& 1.22 	& -4\\
277  &	3:42:22.76	-35:09:15.0	& 12.65	& 17.60	& 13.50	& 18.63	& 11.52	& 16.91	& 0.84 	& 1.14 	& -4\\
279  &	3:42:26.40	-36:41:13.2	& 16.27	& 23.06	& 16.73	& 23.71	& 15.13	& 21.96	& 0.59 	& 1.10 	& -5\\
280  &	3:42:36.96	-35:57:20.9	& 18.24	& 24.13	& 19.17	& 24.71	& 17.29	& 23.24	& 0.75 	& 0.98 	& -5\\
281  &	3:42:38.69	-35:52:03.9	& 17.59	& 23.69	& 18.35	& 24.46	& 16.55	& 22.72	& 0.77 	& 1.07 	& -5\\
284  &	3:42:54.92	-35:20:35.9	& 18.89	& 24.44	& --	& --	& 18.00	& 23.67	& 0.54 	& 1.05 	& -5\\
285  &	3:43:01.95	-36:16:16.4	& 13.78	& 21.72	& 14.24	& 22.20	& 12.90	& 20.86	& 0.41 	& 0.89 	& 7 \\
286  &	3:43:12.69	-34:38:35.0	& 17.68	& 23.06	& 18.23	& 23.90	& 16.58	& 21.90	& 0.71 	& 1.08 	& -5\\
287  &	3:43:13.53	-35:31:07.0	& 17.52	& 23.37	& 18.45	& 24.09	& 16.52	& 22.43	& 0.77 	& 1.06 	& -5\\
288  &	3:43:22.77	-33:56:19.5	& 14.94	& 20.78	& 15.69	& 21.72	& 13.80	& 19.77	& 0.75 	& 1.14 	& -5\\
289  &	3:43:23.20	-34:41:42.6	& 18.10	& 24.07	& 18.68	& 24.55	& 17.15	& 23.27	& 0.65 	& 0.94 	& -5\\
290  &	3:43:37.09	-35:51:17.5	& 11.52	& 19.45	& 12.36	& 20.34	& 10.51	& 18.44	& --	& --	& 5 \\
291  &	3:43:39.72	-35:12:56.6	& 19.26	& 24.29	& --	& --	& --	& --	& --	& --	& -5\\
293  &	3:44:25.19	-35:51:23.4	& 17.10	& 23.02	& 17.96	& 23.80	& 15.84	& 22.32	& 0.72 	& 1.09 	& -5\\
295  &	3:44:29.98	-35:10:41.6	& 18.32	& 23.99	& 18.45	& 24.53	& 17.03	& 23.26	& 0.45 	& 0.98 	& -5\\
296  &	3:44:32.94	-35:11:44.8	& 15.75	& 21.60	& 16.46	& 22.53	& 14.71	& 20.62	& 0.70 	& 1.06 	& -5\\
297  &	3:44:39.28	-35:58:57.2	& --	& --	& --	& --	& 16.44	& 23.12	& --	& --	& -5\\
298  &	3:44:44.41	-35:41:00.5	& 16.02	& 21.39	& 16.76	& 22.32	& 14.92	& 20.55	& 0.74 	& 1.08 	& -5\\
299  &	3:44:58.67	-36:53:40.4	& 16.42	& 22.44	& 16.95	& 22.97	& 15.42	& 21.57	& 0.51 	& 0.10 	& 7 \\
300  &	3:44:59.94	-36:19:09.5	& 15.38	& 22.56	& 16.23	& 23.40	& 14.29	& 21.45	& 0.81 	& 1.13 	& -5\\
301  &	3:45:03.65	-35:58:21.7	& 13.19	& 18.66	& 14.04	& 19.68	& 12.01	& 17.66	& 0.83 	& 1.18 	& -4\\
302  &	3:45:12.32	-35:34:14.2	& 15.46	& 21.83	& 15.59	& 21.99	& 14.89	& 21.92	& --	& --	& 8 \\
303  &	3:45:14.08	-36:56:12.4	& 15.06	& 21.20	& 15.84	& 22.09	& 13.99	& 20.29	& 0.76 	& 1.11 	& -5\\
304  &	3:45:30.83	-34:30:18.3	& 17.97	& 24.50	& 19.51	& 24.56	& 17.25	& 23.57	& --	& --	& -5\\
306  &	3:45:45.41	-36:20:45.2	& 15.66	& 21.16	& 16.04	& 21.51	& 14.78	& 20.55	& --	& --	& 9 \\
307  &	3:45:47.80	-35:03:37.3	& 17.08	& 24.33	& 17.81	& 24.45	& 15.93	& 22.22	& 0.69 	& 0.10 	& -5\\
308  &	3:45:54.87	-36:21:29.8	& 13.10	& 20.66	& 13.90	& 21.42	& 11.91	& 19.56	& 0.72 	& 1.18 	& 7 \\
309  &	3:46:08.24	-36:49:21.7	& 18.45	& 24.47	& --	& --	& 16.68	& 23.60	& --	& --	& -5\\
310  &	3:46:13.82	-36:41:48.0	& 12.60	& 19.18	& 13.46	& 20.19	& 11.36	& 18.13	& 0.86 	& 1.21 	& -1\\
312  &	3:46:19.01	-34:56:36.2	& 11.74	& 19.66	& 12.46	& 20.51	& 10.59	& 18.45	& 0.71 	& 1.16 	& 5 \\
313  &	3:46:33.45	-34:41:09.1	& 16.84	& 22.74	& 17.64	& 23.36	& 15.75	& 21.73	& 0.72 	& 1.11 	& -5\\
316  &	3:47:01.52	-36:26:14.9	& 15.82	& 22.51	& 16.80	& 23.38	& 14.75	& 21.56	& 0.87 	& 1.08 	& -5\\
318  &	3:47:08.17	-36:19:36.3	& 15.54	& 22.45	& 16.41	& 23.31	& 14.60	& 21.49	& 0.79 	& 1.04 	& -5\\
\hline
\label{tab_resb}
\end{tabular}\\
\small $^1$ Colours are in the Cousins system.
}
\end{table*}

\subsection{Selection limits}

\begin{figure}
\begin{center}
\includegraphics{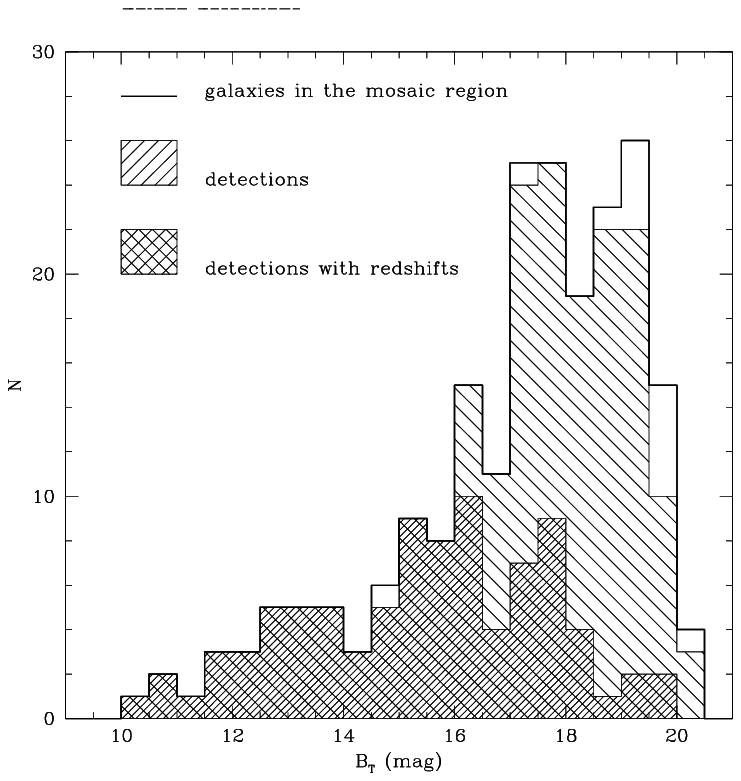}
\caption{Distribution of $B_T$ magnitudes (FCC) of catalogue members
and SExtractor detections. Solid line represents galaxy members in our
catalogue of the central \4 region. The hatched histogram indicates
the galaxies detected. The cross-hatched
histogram indicates the proportion of galaxies with radial
velocities.}
\label{sample}
\end{center}
\end{figure}

Figure \ref{sample} shows the distributions of B$_{T}$ magnitudes from
the FCC for the cluster members in our revised catalogue. The sample
of cluster members with redshifts is limited to a magnitude of
\bj\-19.8, the detection limit of the FCSS survey. Below this
magnitude we assume Ferguson's membership classifications hold.

Of the 213 cluster members in our revised catalogue, 92 (43\%)
galaxies have redshifts confirming them to be cluster members. 200 of
the 213 galaxies (94\%) were detected with SExtractor. Since our mosaic
overlaps the fields from the FCSS, 44\% of the galaxies detected (88)
are confirmed members. Galaxies which were contaminated by foreground
stars or suffered overcrowding have been omitted. Generally these were
the faintest galaxies in the catalogue. Our final sample consists of
photometry of 190 galaxies of which 85 (45\%) are confirmed
members. Galaxies were detected over the entire magnitude range of the
catalogue to a measured apparent V magnitude of 20.5 and peak surface
brightness of 24.9 (V mag  arcsec$^{-2}$).

\section{Surface brightness - magnitude relation}
It has been well established \citep{Ferguson1988,Ferguson1994} that
there exists an empirical surface brightness-magnitude relation for
normal elliptical and dwarf galaxy populations. The physical basis of this
relation is not well understood and many authors have suggested that it is
simply a manifestation of the methods in which galaxy samples are
selected \citep{Phillipps1987,Phillipps1988}.

In this section we discuss the surface brightness-magnitude relation
and its use for determining cluster membership. We also present the
relation for our sample of Fornax Cluster galaxies, paying particular
attention to the location of the ultra-compact dwarf galaxies. Our
surface brightness relation spans a luminosity range over four
magnitudes, to a limit of \bj \- 19.8.

Unlike previous studies which have concentrated on samples of a given
morphological type \citep{Binggeli1991,Bothun1991}, our cluster sample
includes a wide variety of morphological types, observed with the same
instrument and measured using the same technique.

\subsection{Cluster membership}

The greatest problem concerning the study of the faint cluster
population is distinguishing members from the background
population. Few clusters have known radial velocities of their
faintest members. The 2dF and Flair-II Fornax Cluster surveys provide
the largest sample of confirmed faint members. In this section we
investigate the surface brightness-magnitude relation for members and
non-members and its application to cases where radial velocities may be
unobtainable.

\begin{figure*}            
\centering                                                             
\includegraphics{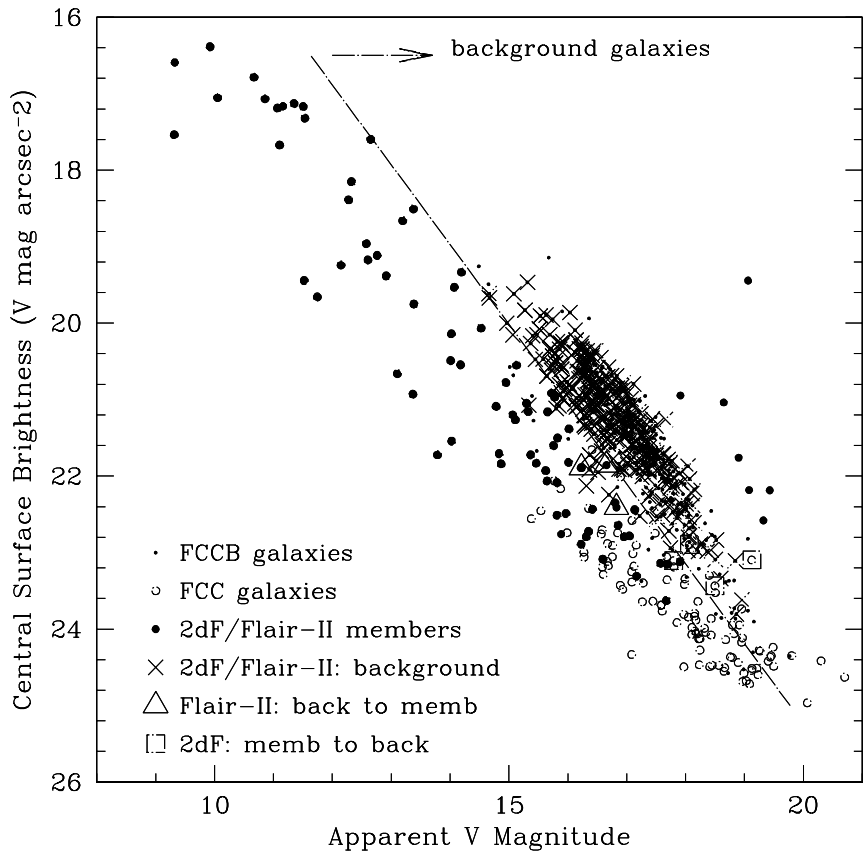}
\caption{Surface brightness-magnitude diagram for cluster members and
background galaxies. Cluster members (from FCC): unconfirmed (open
circles), confirmed by Flair-II/2dF (filled circles). Background
galaxies (from FCCB): unconfirmed (points), confirmed by Flair-II
(crosses). Triangles are Flair-II confirmed members originally
classified as background galaxies. Squares are 2dF confirmed
background galaxies originally classified as cluster members. The
dashed line represents the segregation between populations (SB = 1.06 V
+ 4.16)}
\label{fig_SBmembership}
\end{figure*}

Figure \ref{fig_SBmembership} is the surface brightness relation for
Fornax Cluster members and background galaxies in the region defined
by our mosaic. Unconfirmed background galaxies from the FCC are
plotted as points and unconfirmed members as circles. Confirmed
cluster members with redshifts determined from the FCSS and Flair-II
surveys are plotted as filled circles and background galaxies as
crosses.

Members and background galaxies exhibit strong surface brightness
relations, background galaxies having a higher central surface
brightness for a given magnitude. The dashed line separates the two
populations; member galaxies occupy the region to the left.
This separation between the relations has been
optimised such that the minimum number of confirmed background
galaxies would be classified as cluster members. Of the 224 galaxies
to the left of the dashed line only 6\% are confirmed background
galaxies. 

We detected three background galaxies from the FCC which from the
Flair-II survey, were found to be cluster members. Their location on
the surface brightness magnitude plot places them clearly within the
cluster member population. Similarly, we detected all four of the new
background galaxies, confirmed by the FCSS observations. Three of
these lie well within the population of background galaxies. 

For surface brightnesses \<21.5 \magsqsec the separation between
populations is obvious however as surface brightness increases
distinguishing between background and member galaxies becomes more
difficult. In this regime a higher proportion of background galaxies
may masquerade as cluster members and vice versa. The inability of the
surface brightness relation to characterise the faintest cluster
members is entirely due to the resolution and seeing of the data. This
effect and its implications will be left to the discussion.

The small sample of galaxies to the right of the background sample
are the ultra-compact dwarf galaxies (UCDs). These high surface-brightness
galaxies occupy the the intermediate region between the cluster dwarf
population and galactic globular clusters \citep{Ferguson1994}. 
In this case membership judgements based on surface brightness and
luminosity break down and radial velocities are needed to determine
membership.

\subsection{Surface brightness-magnitude relation for cluster members}
\begin{figure*}
\begin{center}
\vspace{-8cm}
\includegraphics[scale=1.5]{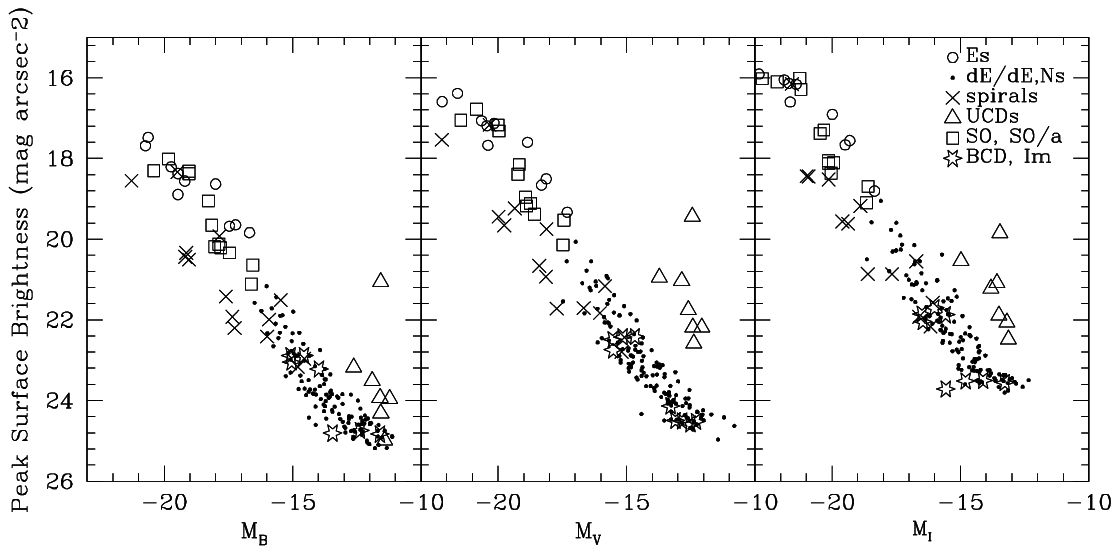}
\caption{Magnitude-surface brightness diagram for Fornax Cluster
members.  A discontinuity between the E and dE populations is observed
in the B band relation.  }
\label{fig_SBfnxgals}
\end{center}
\end{figure*}

Our sample of 190 member galaxies covers a wide range of morphological
types. Figure \ref{fig_SBfnxgals} shows the surface brightness
relation for the entire sample in all bands. The surface brightness
relation for Fornax galaxies follows a narrow locus and as expected
late-type galaxies tend to have a lower surface brightness for a given
magnitude.

The UCDs lie in an isolated region in the surface brightness-magnitude
diagram suggesting they form a distinct class of object. Their
luminosities overlap the faint end of the dwarf population,
-13\<\mb\<-11, based on a distance modulus of 31.5
\citep{Drink2001ApJ}.  However lower limits of their surface
brightness (\- 22.4 \magsqsec) suggest core luminosities much brighter
than any other Fornax Cluster dwarf of the same magnitude. They are
also brighter than any of the globular clusters associated with NGC
1399, the most luminous globular having an absolute magnitude of \mb
\- -11 \citep{Forbes1998}.
 
The observed gap between the E and dE galaxies at an absolute magnitude
of B \- -17 is consistent with the results of \cite{Ferguson1994}, The
dichotomy is progressively less pronounced as you move to redder
wavelengths. The V and I-band relations show a fairly smooth
transition between the dwarf and elliptical populations. A similar
result has been recently obtained by \cite{Graham2003} from HST
photometry of dE galaxies and bright E galaxies in Coma.

The population of elliptical galaxies follow a fairly well defined
sequence with central surface brightness increasing with decreasing
luminosity, with similar scatter as the dE galaxies. This is the opposite
trend to the results of \cite{Kormendy1985,Graham2003}, and we suspect it is due
to the fact that we are not resolving the cores of the galaxies. 
This will be discussed in more detail in section 7.

\section{Colour analysis of cluster galaxies}
\begin{figure}
\begin{center}
\includegraphics{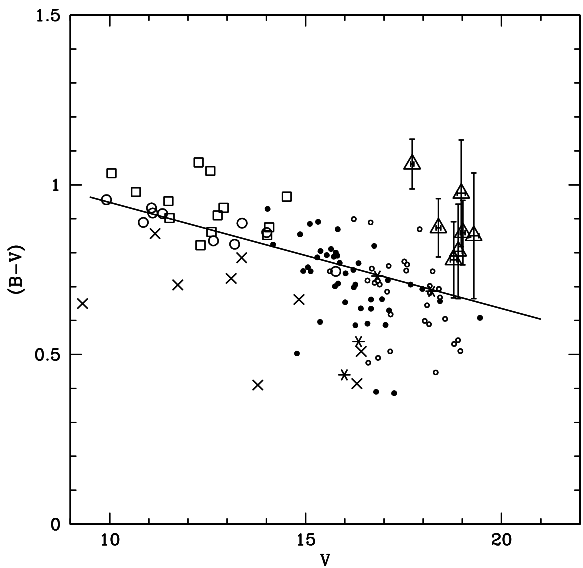}
\caption{Colour-magnitude plot of cluster members; circles-Es,
   boxes-SO/a, small circles-dEs, filled circles-dE,Ns from the FCC,
   crosses-spirals, asterix-BCD/Im and triangles-UCDs. Colours
   are in the Cousins system. The solid line represents a least-squares 
fit to the ellipticals.}
\label{fig_colmag}
\end{center}
\end{figure}

We have determined B-V and V-I colours for 113 cluster
galaxies. Reddening is very small towards Fornax ($E_{(B-V)}
\simeq0.01\;\mbox{mag}$, \cite{Sch1998}) and can be neglected in
interpreting the results. The colour-magnitude plot for
the entire population is shown in figure \ref{fig_colmag}. Again we
adopt the ``t-type'' classification scheme described above.

The colour-magnitude properties of galaxies are an important aspect of
galaxy evolution studies. The well known colour-magnitude relation for
bright cluster ellipticals
\citep{Terlevich2001,Visvanathan1977,Vazdekis2001} is generally
attributed to a metallicity effect. More massive galaxies tend to have
higher metallicities as a result of their large binding energies,
appearing redder than the less massive galaxies.

Our data show that early-type galaxies become progressively redder
with increasing luminosity. Galaxies classified as SO/a morphological
types also follow the same colour-magnitude relation as elliptical
galaxies.  A least squares fit to the E and SO population gives a
slope of -0.034 $\pm$ 0.006. This is consistent with the observations
of \cite{Griersmith1982} who find a slope of -0.038 $\pm$ 0.005 for
both Fornax and Virgo elliptical and SO populations. The V-I colour
magnitude relation gives a similar result. A least squares fit to the
data yields: (V-I) = -0.028V + 1.52 with an RMS of
0.8. \cite{Hilker2003} find a similar result for their study of dwarf
spheroidals in Fornax.

The relation for early-type galaxies is fairly tight which is
expected for bright cluster galaxies.  In contrast, luminous late-type
galaxies show a much broader distribution and as expected are
significantly more blue. This is a similar result to that found in
other clusters, for example in Coma \citep{Terlevich2001} and reflects
the different star-formation histories.

The colour-magnitude properties of dwarf galaxies in the Fornax
Cluster have been discussed by many authors
\citep{Evans1990,Phillipps1987}. The dE and dE,N populations follow a
colour-magnitude relation in the sense that the brighter dE and dE,N galaxies
are redder. A similar result was also obtained by
\cite{Caldwell1987} and \cite{Hilker1999:I}. Our results are
consistent with previous studies and we find that both dE and dE,N galaxies
follow the same colour-magnitude relation. As expected the scatter
increases towards the faint end of the population and blueward of the
relation for early-type galaxies.

By contrast, the UCD galaxies lie significantly above the
colour-magnitude relation for early-type galaxies although the amount
of scatter redward of the relation for ellipticals is comparable to the
bluest dwarfs.  Previous photometric studies of dE,N galaxies suggest that
there is no measurable difference between the colours of their cores
and halos \citep{Caldwell1987}.  With the exception of the brightest,
the location of the UCDs on the colour-magnitude is therefore
consistent with the galaxy threshing model of a bright dE,N galaxies
\citep{Bekki2001}. As the halo of the dE,N is stripped and the
luminosity decreases the colour of the galaxy core remains the
same. We would expect the remnant nuclei, UCD to be shifted off the
locus of the the colour-magnitude relation.

\begin{figure}
\begin{center}
\includegraphics{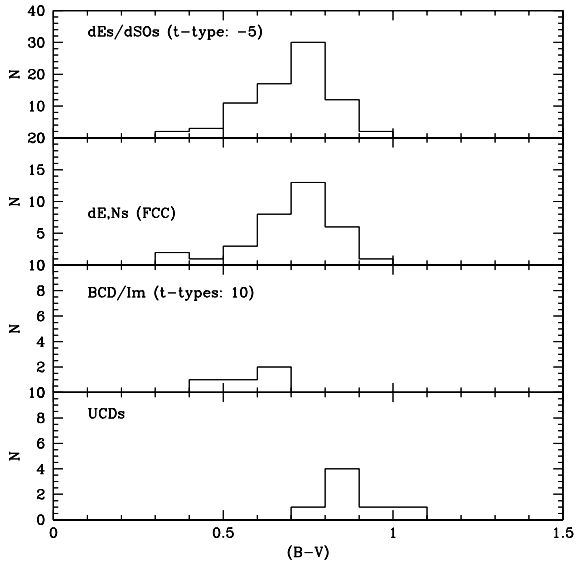}
\caption{Histograms of the B-V colours of Fornax dwarf; dE,Ns
classified by Ferguson have been separated from the dE population based
on the t-type scheme. }
\label{fig_hist}
\end{center}
\end{figure}

The histograms in Figure \ref{fig_hist} show the colour distribution
of the dwarf population, separated by their morphologies or
``t-types''. The dwarf elliptical population has been further divided
into nucleated and non-nucleated using the original classification
by Ferguson \citep{Ferguson1989}.

\begin{table}
\caption{Mean B-V \& V-I colours of the dwarf population}
\begin{tabular}{llll}
\hline
Sample & n	& $\langle$B-V$\rangle$, $\sigma$ & $\langle$V-I$\rangle$, $\sigma$  \\	
\hline
dEs/dSOs (t-types: -5) & 76	& 	0.70,  0.12  &	1.06, 0.08 \\
dE,Ns (FCC)            & 34	&	0.71,  0.13  &	1.06, 0.09\\
BCD/Im (t-types: 10)   & 4	&	0.57,  0.10  &	1.02, 0.10\\
UCDs                   & 7 	&	0.89,  0.12  & 	1.09, 0.14\\
\hline
\label{tab_mean}
\end{tabular}\\
\small Note: Colours are in the Cousins system.
\end{table}

The mean colours of each population are also given in Table
\ref{tab_mean}.  Using {\em KS-} and {\em t-tests} we find that the
distribution and mean B-V colours of the UCDs are significantly
different from the 'normal' dE population, at the 99\% confidence
level. The UCDs are redder than the dE and dE,N
($\langle$B-V$\rangle$=0.7) population having a $\langle$B-V$\rangle$
colour of 0.89. Their V-I colours, $\langle$V-I$\rangle$=1.09, are
consistent with the colours of galactic globular clusters of the
Harris Catalogue \citep{Harris1996}. They are also typical of the
globular clusters of NGC 1399 \citep{Kissler1997,Dirsch2003}.  Our results
are consistent with the most recent published colours of
\cite{Mieske2002}.

\cite{Bothun1991} also investigated the colours of a small sample of
LSB Fornax galaxies. Their sample of dE galaxies were found to be relatively
blue with a $\langle$B-V$\rangle$ of 0.6. Since there is a trend for
photographically selected LSB galaxies to be blue \citep{Impey1988}
their result was interpreted to be partially due to selection effects.
The mean for our sample of dwarf galaxies is ($\langle$B-V$\rangle$=
0.69) which is consistent within errors to the results of
\cite{Bothun1991}.  

As expected, late-type dwarfs are bluer ($\langle$B-V$\rangle$= 0.57)
than early type dwarfs, reflecting the relative
star formation histories of the populations.

\section{Discussion}
\subsection{Membership}
Previous photographic studies of galaxy clusters have relied on a
multitude of techniques used to distinguish cluster members from
background galaxies. These include statistical methods based on colour
selection, morphology and surface brightness measurements and 
in most cases this has proved to be adequate for the study of cluster
populations.

In a study of the Virgo cluster, \cite{Sandage1984} concentrated on
the more complicated identification of low-luminosity dwarf
galaxies. These were judged to be cluster members either from their
uniquely low surface brightnesses and/or morphological
classifications. The Fornax Cluster Catalogue \citep{Ferguson1989} was
also constructed using the same technique. From the results of more
recent spectroscopic surveys of the Fornax Cluster only a small percentage
of galaxies in the FCC were assigned incorrect membership classifications.

The motivation of our investigation was to test whether the empirical
surface brightness relation is simply a result of the way galaxy
samples are chosen. Our sample is based on the results of the Fornax
Cluster Spectroscopic Survey and the Flair-II Fornax survey in order
to obtain as many cluster and background galaxies with confirmed
redshifts. We have assumed the original Ferguson classifications of
the fainter dwarf population hold. In the future we hope measure
redshifts for these galaxies to confirm cluster membership to fainter
limits.

The confirmed cluster members and non-members occupy two distinct
regions on the surface brightness-magnitude plot. We have found that
both confirmed and unconfirmed background galaxies lie along the same
locus and have a higher surface brightness than cluster members of the
same magnitude. For surface brightnesses fainter than 22 V \magsqsec
the two relations merge and distinguishing between members and
non-members requires additional radial velocity measurements in order
to make judgements of cluster membership.

The main limitation in the interpretation of the results is the
resolution of the data. Our peak surface brightness measurements are
largely influenced by resolution and seeing. The relation for
background galaxies is simply a manifestation of this effect.  Seeing
acts in such away that the measured peak surface brightness will
always be fainter than the real central surface brightness. As
background galaxies are less well resolved than cluster galaxies the
seeing will have the dramatic effect of suppressing the central
surface brightness. In particular background galaxies containing small
cores, such as ellipticals and small scale-size dwarfs will be
seriously affected by this method. With higher resolution observations and better
seeing the separation between cluster and background galaxies would be
more obvious down to much fainter surface brightness limits. 

\subsection{Surface brightness-magnitude relation}
Our surface brightness-magnitude relation for Fornax Cluster members
is consistent with previous studies
\citep{Phillipps1987,Caldwell1987}. The E and dE galaxies follow a
well defined sequence with increasing surface brightness corresponding
to an increase in galaxy luminosity.

For our large sample of dwarf galaxies the observations are consistent
with the recently published results of \cite{Deady2002} who
investigated surface-brightness magnitude relation for 24 Fornax
Cluster dwarfs. Total magnitudes were determined using the same
methods as the photometry we have presented in this paper. However
extrapolated central surface brightnesses were obtained after fitting
the galaxies with Sersic profiles.

The canonical surface brightness relation of \cite{Ferguson1994} shows
a clear break between the two populations.  In that study the dE
galaxies were plotted using photometric data from a complete sample of
$\approx$200 early type galaxies in the Virgo cluster where the `mean'
central surface brightness was obtained using King models
\citep{Binggeli1991}. The E galaxies and bulges were data obtained by
\cite{Kormendy1985} and included as many galaxies as possible which
were near enough to have their cores resolved.  Their results show
that dwarf galaxies occupy a region of the plot with absolute
magnitude limits of -16 \< \mb \< -8, the relationship indicating an
increasing mean surface brightness with increasing luminosity. A break
between the E and dE populations was identified.

Our model independent results are consistent with this observation.
This break is clearly seen in the B band relation at a surface
brightness of 21 B \magsqsec and becomes progressively weaker as you
go to longer wavelengths.  A discontinuity in surface brightness at \-
21 B \magsqsec is not entirely unexpected since it is known to be a
transition point where many other properties within the E family are
changing \citep{Jerjen1997}.

In the canonical surface brightness relation the relationship for E
galaxies is somewhat different to the dE population, the mean surface
brightness increasing with decreasing luminosity. The interpretation
of this segregation was concluded to result from the disparity of the
model profiles fitted to each morphological type \citep{Binggeli1991}.
Our model independent surface-brightness relation for E galaxies
opposes this trend of increasing surface brightness with increasing
luminosity. Our relation for E galaxies is similar to the results of
\cite{Jerjen1997}, who also find that the population of dE and E
galaxies smoothly and continuously merge. Their analysis was based on
fitting Sersic profiles to the overall shape of galaxies, ignoring the
innermost 3'' ($\sim$ 300pc) of the profiles. Their extrapolated
surface brightnesses were systematically higher than the central surface
brightnesses observed by \cite{Kormendy1985} resulting in a shift that
resolves the E-dE dichotomy. From this analysis is was concluded that
the dichotomy was not evident because the cores of the E galaxies were
not resolved.

A more recent analysis by \cite{Graham2003} show that the alleged
dichotomy between E and dE galaxies can be resolved using Sersic
profiles and when necessary a central point-source or PSF-convolved
Gaussian. The relation for E galaxies depends on the ability to
resolve the cores of the brightest elliptical galaxies. The dE
galaxies are shown to display a continuous sequence with the brighter
E galaxies such that the central surface brightness increases with
increasing magnitude until core formation causes the most luminous E
galaxies to deviate from the relation \citep{Graham2003}.  Since our
data is limited by resolution (2.025'' pix $\approx$ 200pc) we do not
observe the relation for the most luminous E of increasing surface
brightness with decreasing luminosity.

To test this hypothesis we have obtained high-resolution multicolour
imaging of the cluster using the CTIO 4m Blanco Mosaic telescope.  The
image resolution (0.27'' pix $\approx$ 25pc) is such that we will be
able to resolve the cores of the E galaxies and see the trend of
increasing surface brightness with increasing luminosity.

\subsection{Dwarf Galaxies and the origin of the UCDs}
It is not surprising that we find a spread of colours for the dwarf
population since their low binding energies mean they are very
susceptible to the cluster environment and can therefore display a
wide range of evolutionary histories, even within the same subclass
\citep{Grebel2001}. Using integrated galaxy colours to investigate
dwarf galaxy evolution is problematic. Galaxy colours result from many
factors including galaxy metallicity, dust content and star-formation
histories. For their sample of ten Fornax dE,Ns \cite{Held1994} found
a range of metallicities similar to those of intermediate
([Fe/H]\--1.4 dex) to metal rich ([Fe/H]\--0.7 dex) globular
clusters. These metallicities are well correlated with the colours of
\cite{Caldwell1987} which are consistent with the colours from our
analysis.

Our photometry of the ultra-compact dwarfs exemplifies the problem of
determining the membership status of high surface brightness
objects. These objects were previously classified as foreground
stars. Their isolated location on the surface brightness-magnitude
plot seems to indicate they form a distinct class of objects.

From our results it is likely that we are observing remnant nuclei of
a small sample of nucleated dwarf ellipticals which have been tidally
stripped through their interaction with the central cD galaxy NGC
1399. This ``galaxy threshing'' is supported by simulations
\citep{Bekki2001} of infalling nucleated dwarf elliptical galaxies,
which are tidally stripped as they orbit a central cD galaxy. The
compact nucleus is weakly influenced by tidal forces. Since the
contribution of light from the nuclei is typically 2\% for nucleated
dwarfs \citep{Binggeli1991}, the central surface brightness remains
the same as the galaxy halo is stripped, however the apparent
magnitude increases. On a surface brightness-magnitude plot this
``stripping'' manifests as a shift off the locus of the dE population
towards a much fainter apparent magnitude. A further test for this
theory would be to search the cores of clusters such as Virgo and Coma
where similar objects may have been previously overlooked. Their
location on the cluster surface brightness relation and colours would
provides additional selection information.

\begin{figure}                   
\centering                                                             
\includegraphics{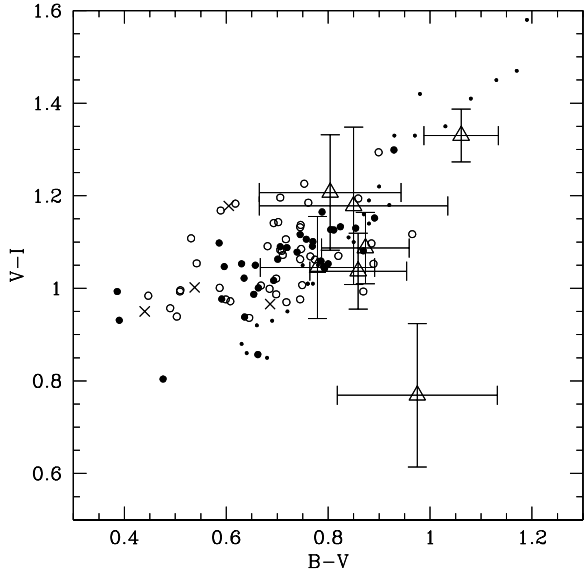}
\caption{Colour-colour plot of the cluster dwarf population; dEs
   (small circles), dE,Ns from the FCC (filled circles), BCD/Im (small
   crosses) and UCDs (triangles) with error bars. Galactic globular
   clusters from the Harris catalogue \citep{Harris1996} are shown as
   small points.}
\label{fig_col}
\end{figure} 

It has also been suggested that one or more of the UCDs may form part
of the bright end of the globular cluster population associated with
NGC 1399 \citep{Hilker1999:I,Mieske2002}. Figure \ref{fig_col} is a
colour-colour plot of the cluster dwarf population. The UCDs lie well
within the cluster dwarf population and have similar colours to
galactic globular clusters \citep{Harris1996}. UCD5 lies away from the
remaining population.  As noted in section 4 we suspect higher
resolution measurements will yield a similar V-I colour to the
remaining population.

The errors of the photometry make an interpretation of the UCD
colours by single burst stellar population models
(i.e.\cite{Bruzual1993,Worthey1994}) unreasonable.  Again higher
resolution observations will be required to make useful conclusions.

We are currently analysing deep multicolour imaging of the central 1
deg$^2$ region of the cluster, taken with the CTIO 4m Blanco Mosaic
telescope. The image resolution of the Mosaic data (0.27''/pixel) is
such that the errors will be considerably less than the current
photometry. Using the Sloan filter set (u'g'r'i'z') we hope to
accurately determine colours of the cores and halos of the dE,N
population and the UCDs as well as the globular cluster population
associated with NGC 1399. This will enable us to test the two main
hypotheses for the origin of the UCDs and so we defer any further
discussion of their colours until the presentation of the Mosaic data.

\section{Conclusions}
We have presented BVI photometry for a sample of 190 Fornax Cluster
members. Our revised Fornax Cluster catalogue incorporates results
from a number of recent spectroscopic surveys and also includes a
small sample of recently discovered ultra-compact dwarf galaxies.

The UCDs have absolute magnitudes in the range -13 \< \mb \<
-11. These results are consistent with previous R-band photometry of
photographic images of the UCDs \citep{Drink2000} and B photographic
colours of \cite{Deady2002}. Although their colours
$\langle$V-I$\rangle$=1.09 and $\langle$B-V$\rangle$=0.89 suggest an
older stellar population not unlike globular clusters, they are much
brighter than the most luminous of the globulars (\mb \- -11)
associated with NGC 1399 \citep{Forbes1998}. The UCDs lie off the
surface-brightness magnitude correlation for dE galaxies which is
consistent with the ``galaxy-threshing'' scenario by \cite{Bekki2001}.
This hypothesis is also supported by their location off the locus of
the colour-magnitude relation for dE,Ns.

We have investigated the surface-brightness magnitude relation for
confirmed cluster members and background galaxies. Both populations
occupy two distinct regions on the surface-brightness magnitude plot
however at low luminosities the relations merge. This is due to the
fact that the relation for background galaxies is largely influenced
by the resolution and seeing of the observations.

For a more comprehensive analysis of the surface brightness-magnitude
relation and the population of ultra-compact dwarfs we require much
higher resolution observations.  We are currently analysing deep
multicolour imaging of the central 1\deg region of the cluster taken
with the CTIO 4m Blanco telescope. The image resolution of
0.27''/pixel will provide us with more accurate surface brightness and
colour measurements. Radial surface brightness profiles of the
infalling dwarf population and the UCDs will further aid the
investigation into the origin of the ultra-compact dwarfs. In
addition, a comparison of the colours of cores and halos of infalling
dE,N galaxies and the globular clusters associated NGC 1399 will
enable us to further constrain the origin of the UCDs.

We also plan to investigate the population of infalling dwarf galaxies
which lie outside our mosaic area, with particular emphasis on the
merging Fornax sub-cluster. This will enable us to make a more
detailed study of the colours of the Fornax population and the effect
of the cluster environment on galaxy evolution.

\section{Acknowledgements}
We wish to thank Bryn Jones for helpful discussions on the photometry
and suggestions regarding the analysis of our data.  We are grateful
to Mike Fall for suggesting the colour-magnitude analysis of the
UCDs. We thank the referee M. Hilker for his very useful comments
which improved the paper. We also thank our colleagues from the Fornax
Cluster Spectroscopic Survey for providing the velocity data for our
sample. This project was supported by grants from the Australian
Research Council and the Australian Nuclear Science and Technology
Organisation Access to Large Research Facilities scheme. This material
is based in part upon work supported by the National Science
Foundation under Grant No.~9970884 and carried out at the Institute of
Geophysics and Planetary Physics, under the auspices of the
U.S. Department of Energy by Lawrence Livermore National Laboratory
under contract No.~W-7405-Eng-48.

\bibliographystyle{mn2e}
\bibliography{mn-jour,references}
\end{document}